\documentclass[aps,prl,twocolumn,groupedaddress,showpacs,floatfix]{revtex4}

\usepackage{graphicx}
\usepackage{natbib}
\usepackage{amsmath}

\begin{document}

\title{Trapping of ultracold polar molecules with a Thin Wire Electrostatic Trap}

\author{J. Kleinert}
\author{C. Haimberger}
\author{P. J. Zabawa}
\author{N. P. Bigelow}

\affiliation{Department of Physics and Astronomy, and The
Laboratory for Laser Energetics\\ The University of Rochester,
Rochester, NY 14627}

\date{\today}

\begin{abstract}
We describe the realization of a DC electric field trap for
ultracold polar molecules, the Thin WIre electroStatic Trap (TWIST). 
The thin wires that form the electrodes of the TWIST allow us to superimpose the trap onto a magneto-optical trap (MOT). In our experiment, ultracold polar NaCs molecules in their electronic ground state are created in the MOT via photoassociation, achieving a continuous accumulation in the TWIST of molecules in low-field seeking states. Initial measurements show that the TWIST trap lifetime is limited only by the background pressure in the chamber. 
\end{abstract}


\maketitle 

Recently, there has been a growing interest in the study of ultracold molecules. While the rich internal structure of molecules poses new challenges in terms of the methods used for cooling and trapping them,
the possible applications of cold molecules are remarkable.  This is especially true for strongly polar molecules\cite{Bohn1,Wieman1,Hinds1,Zoller1,Zoller2,DeMille1}. Examples include experiments in superchemistry, tests of fundamental symmetries such as the search for a permanent dipole moment of the electron, quantum information processing and the realization of dipolar quantum gases such as polar Bose-Einstein Condensates.
For most of these applications, both rovibrational state selectivity and robust trapping schemes are essential. 
Multiple approaches have been pursued to achieve these two goals simultaneously. For example, buffer gas cooling joined with magnetic trapping \cite{Doyle2}, Stark deceleration \cite{Meijer1} and velocity selection \cite{Rempe1}
joined with electrostatic/electrodynamic (DC/AC) trapping
\cite{Meijer2,Meijer4,Rempe2} and photoassociation
\cite{Julienne2,Marcassa1,Stwalley1,DeMille2,Bigelow2,Weidemueller1} joined with
magnetic trapping \cite{Stwalley2} have all been realized and have attracted substantial interest. Each technique has its advantages as well as limitations,
often involving trade-offs between state-selectivity and trapped molecule number, density and temperature.

Photoassociation offers the unique starting point in terms of the formation of cold polar molecules because it transfers the initial ultracold temperature of the constituent atoms onto the molecules.
This easily results in molecular kinetic temperatures of a few 100 $\mu$K and with only a little extra effort, much lower temperatures can be achieved by cooling the atoms beyond the Doppler limit with well
established methods \cite{Weiner1}. In terms of state-selectivity and control, pump-and-dump and STIRAP (STImulated Raman Adiabatic Passage) based methods have both been demonstrated recently \cite{DeMille5,Grimm3}. However, to accumulate a sample of ultracold molecules via photoassociation a robust trap is needed that is compatible  with the photoassociation process. 

The preferred starting point  for photoassociation of cold polar molecules are bialkali atomic mixtures, cooled and trapped in a MOT.  Cold dense samples of alkali atoms are readily created and the resulting polar molecules display 
strong dipole moments (typically several Debye) in the deeply bound levels
 of their $X^{1}\Sigma$ state \cite{Dulieu1}. It is important to note, 
however, that magnetic trapping of molecules created in this state is not
a feasible option due to the comparative lack of a magnetic moment in that state. As a result, alternative trapping schemes based on interaction with the molecular electric dipole moment (EDM) have been considered.  In these approaches, polar molecules in their absolute ground state (electronic as well as ro-vibrational) have already been successfully trapped in DC/AC electric-field based traps \cite{Rempe2,Meijer4}.   However, to date this approach has only been demonstrated with molecules cooled using methods other than photoassociation and at temperatures in the range of 1-500mK.
Using a very different approach, several groups have also proposed trapping schemes based on radiative interactions such as the optical dipole traps \cite{Knize1} as well as a microwave \cite{DeMille4} trap. In this letter we demonstrate an attractive solution to both creating and trapping cold polar molecules.  We describe a robust DC electric trap that is superimposed onto a MOT and which takes advantage of the inherent ultracold temperatures and the state selectivity of photoassociation.
\par
Like all bialkali molecules, the electronic ground state of the molecule used in our work, NaCs, consists of a $X^{1}\Sigma$ (``singlet") and a $a^{3}\Sigma$ (``triplet") state. The
EDMs of the deeply bound ($\textsl{v}$=0 -- 30) rovibrational states of
the $X^{1}\Sigma$ potential are much larger (4.6 Debye) than the deeply
bound rovibrational $a^{3}\Sigma$ states ($<$0.1 Debye) and in both cases, highly excited vibrational states close to the dissociation limit generally display vanishingly small dipole moments. We are therefore primarily interested in
deeply bound $X^{1}\Sigma$ states, which cannot be trapped
magnetically
. However, the
strong EDM of these states makes trapping them
via electric fields quite feasible.
\par
Consider the relationships between the molecular states, their quantum numbers, and their EDMs, molecular polarizabilities and Stark effects.
A linear Stark effect in a diatomic molecule can be understood in terms of the response of degenerate (or nearly degenerate) states which are mixed and linearly split by an external electric field - analogous to the Zeemann splitting of the magnetic sub-levels of an atomic hyperfine state. However, such a situation 
only exists in molecular states with a non-vanishing orbital angular momentum along the molecular axis (e.g. a $\Pi$ state).
\par
Therefore, the Stark effect to lowest order in the $X^{1}\Sigma$ state is quadratic:
\begin{equation}
\label{stark}
\Delta E_{Stark} = \frac{d^{2}E^{2}}{B}
\frac{J(J+1)-3 M_{J}^{2}}{2J(J+1)(2J-1)(2J+3)} = \alpha E^{2}
\end{equation}
with d the EDM, E the electric field, B the rotational constant, J the total angular momentum, $M_{J}$ the
projection of the total angular momentum onto the molecular axis and $\alpha$ the molecular polarizability.
\par
The only non vanishing angular momentum of the $X^{1}\Sigma$ state arises from the rotation of the molecule, which is perpendicular to
the molecular axis. Hence, non-rotating $X^{1}\Sigma$ molecules are always high-field seekers (i.e. they experience a negative Stark effect), while 
rotating ones can be low-field seekers (i.e. they experience a positive Stark effect). Also note that the polarizability
decreases quadratically with increasing rotational quantum numbers.
\par
While the EDM is maximal in the lowest vibrational (\textsl{v}=0)  $X^{1}\Sigma$ state (4.6 Debye), it is important to 
recognize that it decreases only a few percent up to \textsl{v}=30 \cite{Dulieu1} because the
molecular potential well is nearly harmonic over this range. For the same reason, the rotational constant decreases only by $16\%$ from the
lowest to the 30th level \cite{comment1}. As the polarizability is proportional to the square of the EDM and
inversely proportional to the rotational constant, these two effects roughly counterbalance each other, causing the polarizability to be
constant within $10\%$ for the lowest 30 vibrational states.
\par
As will be shown elsewhere, one remarkable advantage of the NaCs system is that one can populate states well below
the \textsl{v}=30 level of the $X^{1}\Sigma$ state via single step photoassociation. This fortuitous property of the Na+Cs system
allows us to proceed directly to DC trapping of the photoassociated molecules without going through 
additional state preparation steps such as pump-and-dump or STIRAP based schemes.
\par 
Our trap design hinges on a series of trade-offs that allow us to combine strong electrostatic trapping of molecules with standard MOTs.
 The main feature is the use of a series of small electrodes close to the atom clouds which achieve strong field gradients while keeping the
electrodes sufficiently thin that the light fields of the MOT are only minimally perturbed. When developing the trap, we found that if the wires are chosen to be too thin, they
cannot withstand the electrostatic forces of the field they are creating, thus bend and deform the intended 
field distribution, or even worse they short. We found tungsten wire of 75 $\mu$m diameter the most suitable electrode material.
\begin{figure}[h]
\includegraphics[width=8.5cm]{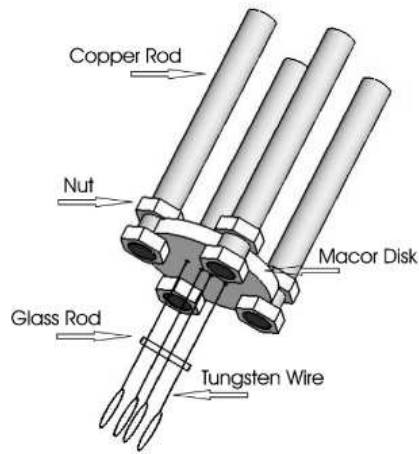}
\caption{\label{fig1} Schematic of the TWIST. 
Each wire is electrically connected to one of the copper rods.}
\end{figure}
We chose 4 concentric rings of 8 mm diameter each, spaced 3, 2 and 3 mm with respect to each other, creating a cylindrical body of
8mm height (Fig. 1).  The exact relative spacings between the rings are maintained by an insulating glass rod attached to the wires just outside the intersecting MOT
light beams.
\par
This electrode configuration can create a static quadrupole field which is suitable for trapping low-field seeking molecules
\cite{Meijer2,Rempe2}. 
The resulting trap volume is calculated to be about 0.065 cm$^3$ which is sensibly matched to the MOT volume.
As the molecules are created inside the TWIST, they are continuously trapped and accumulated.
\begin{figure}[h]
\includegraphics[width=7.5cm]{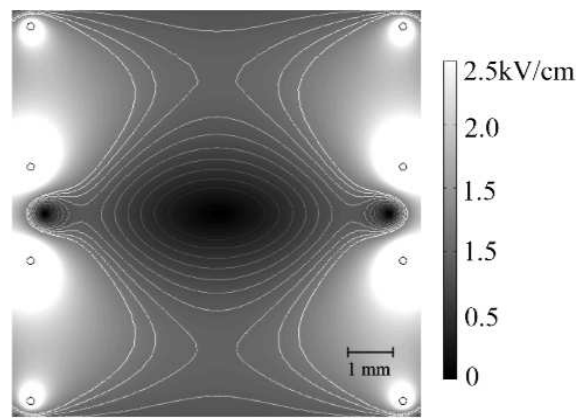}
\caption{\label{fig2} Electric field distribution for the TWIST. The center two electrodes are at +1kV, the outer two are grounded. The diameter of the trapping region is $\approx$5 mm.}
\end{figure}
\par
Given the ultracold temperature of our NaCs molecules ($\approx$ 200 $\mu$K measured by time-of-flight) in conjunction with their large polarizability, only
relatively weak fields are required to trap the low field seeking molecules. For example, charging the inner two rings to +1kV and
grounding the outer two provides a trap depth of $\approx$6 mK for J=1, M$_{J}$ = 0 (\textsl{v}=0-30) states. The resulting field is shown in Fig. 2 
and is our standard trapping configuration. We have found that the electrodes flex only about one wire diameter at a DC electric field of
10kV/cm between any two electrodes \cite{comment2} and hence deflection can be neglected for our first measurements which employ fields less than half as strong.
\par
We determined the lifetime of the trapped molecules in the following way: The inner two electrodes are at +1kV, the outer two are grounded. We start by loading Cs and Na Dark-SPOT-MOTs ($10^7$ Na and $10^7$ Cs atoms, 1.5mm and 2.2 mm diameter, respectively) in the center of the
already charged rings of the TWIST. Excited state molecules are formed via photoassociation by an Argon ion-pumped
ring Ti:Sapphire-laser tuned $\approx$950 GHz below the cesium D2 line on a $\Omega^{\prime}$=2, J$^{\prime}$=4 resonance with a
beam size of 4 mm$^2$ and an intensity of 10W/cm$^2$. 
The J$^{\prime}$=4 state was chosen as it presents the most efficient photoassociation coupling in that particular rotational progression. The molecules
decay into a distribution of rovibrational levels of the electronic
ground state. The molecules that decay into deeply bound vibrational states ($\textsl{v}$=0--30) are confined by the electric
quadrupole field. The trap depths for the populated rotational states are about 670, 390 and 260 $\mu$K for J=3, 4 and 5, respectively. After accumulating molecules for several hundred milliseconds, the MOT beams are shut off via acousto-optic modulators. As the atoms are not confined by the electrostatic trap, 
they leave the detection region within 20 ms due to thermal expansion and gravity; a pure sample of NaCs molecules remains. After a varying
delay time (50-400 ms), the TWIST is switched off and a light pulse of 0.5 mJ energy with a diameter of 1 mm and a wavelength of 593.1 nm ionizes 
the NaCs molecules. The molecular ions are subsequently detected by a channel electron multiplier \cite{Bigelow2}.

\setlength{\unitlength}{1cm}
\begin{figure}[htb]
\begin{picture}(8.5,6.3)(0,0)
\put(0,0){\includegraphics[width=8.5cm]{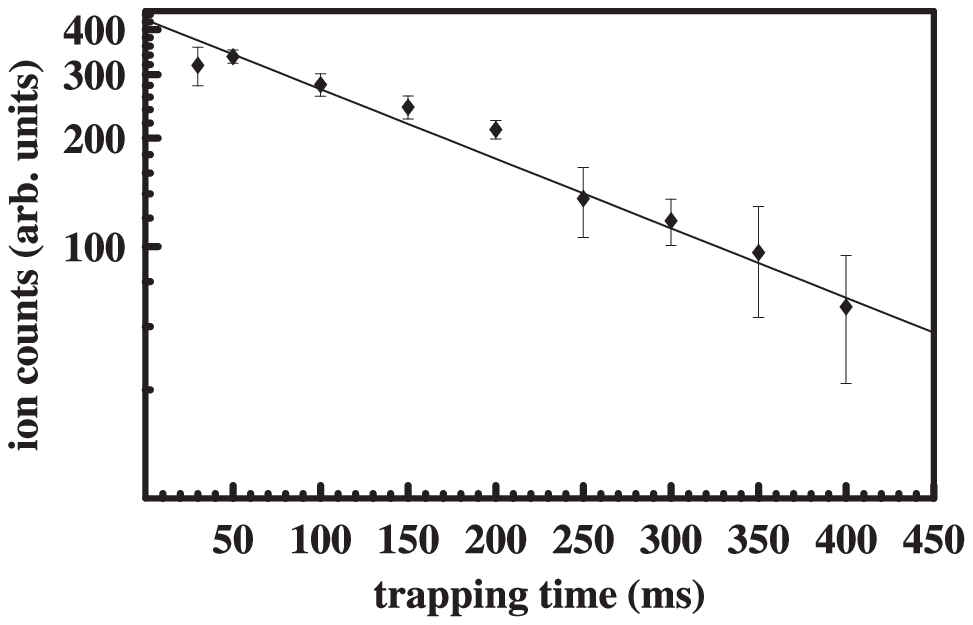}}
\put(1.3,1.3){\includegraphics[width=3.7cm]{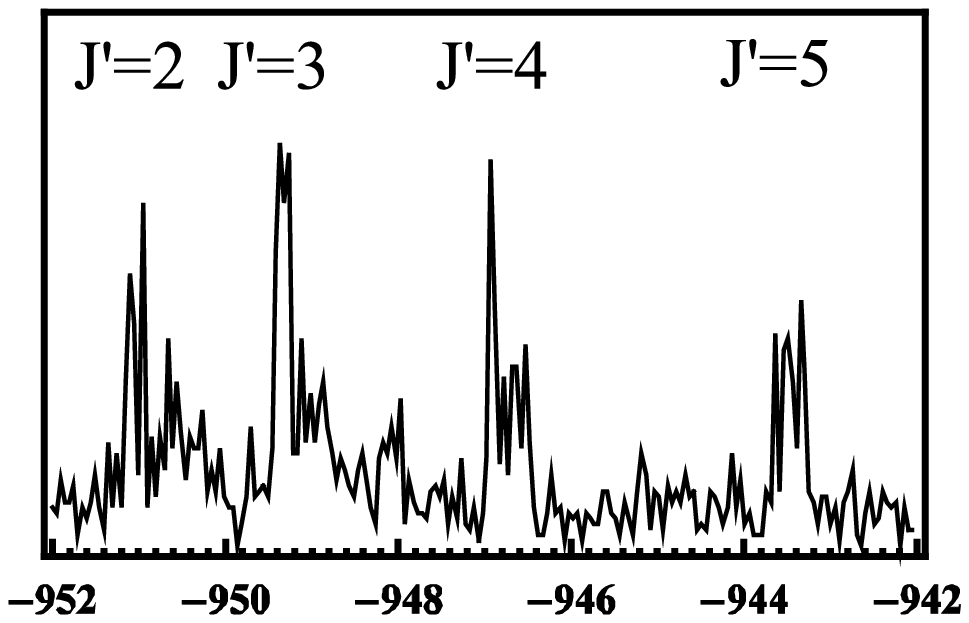}}
  \end{picture}   
\caption{\label{fig3} NaCs molecule signal in the electrostatic trap as a funtion of time. The fitted curve corresponds to a lifetime of 225 ($\pm$ 30) ms of the trapped NaCs. The error bars are the standard deviation derived from three separate measurements at each time step. The inset shows a photoassociation spectrum (-952 to -942 GHz below the Cs 6S$_{1/2} \rightarrow$6P$_{3/2}$ transition frequency)  of molecules trapped for 100ms, demonstrating the ability to trap multiple rotational states.}
\label{addition}
\end{figure}

The data is shown in Fig. 3. The lifetime of the trapped molecules is 225 ms $\pm$ 30 ms, which is comparable to the lifetime of the
trapped atoms in our MOTs (200 ms for Na, 430 ms for Cs). All lifetimes are limited by collisions with background
gas.
To demonstrate the TWIST's ability to trap a variety of rotational states at correspondingly varying trap depths, we also scanned the photoassociating Ti:Sapphire laser across a rotational progression and ionized the trapped molecules after 100 ms. The inset in Fig. 3 shows the resulting spectrum.
\par
Given the ratio of the size of the ionizing light pulse (1 mm
diameter) with respect to the trap size, we estimate the number of trapped molecules in the detected 21st vibrational state to be ~100. This assumes $100\%$
ionization \cite{Bigelow3} and 100 $\%$ detection efficiency and is therefore a conservative lower bound. We estimate the total number of trapped molecules 
to be about a factor of 6 larger, as we populate multiple vibrational states in the range from $\textsl{v}$=19--30 simultaneously via our photoassociation scheme. The
number of trapped molecules is mainly constrained by the quality of our vacuum ($4\cdot10^{-9}$ Torr), which is currently the limiting
factor to the size of the atomic clouds as well as to the lifetime, and hence, the accumulation time of the molecules in the TWIST.
\par
To estimate the impact of an improved vacuum on the trap performance, we solve a rate equation with the following assumptions:
we neglect homonuclear and heteronuclear light assisted collisional losses, as the atom clouds are trapped in a dark-SPOT-MOT and neither of the atom clouds is noticeably depleted in the presence of the other.
The partial pressure of sodium is negligible too, as it is loaded from a Zeemann slower. We estimate our loss rate due to background cesium to be about an order of magnitude smaller than the loss rate due to other background gases at $4\cdot10^{-9}$ Torr. 
Hence, a drop in pressure of an order of magnitude in the chamber would translate into a factor of five more trapped atoms in each of the two atom clouds. As the photoassociation process is not noticeably depleting our atom clouds, this increase will improve the molecular production rate by a factor of  roughly 25. As an improved vacuum also improves the lifetime of trapped molecules in the TWIST, we can expect an additional factor of 5 in the number of trapped molecules due to a longer accumulation time. 
Therefore,  improving the pressure in our experiment by an order of magnitude to  $4\cdot10^{-10}$ Torr will result in $10^4-10^5$ trapped molecules, 
which corresponds to a density of $\sim10^5-10^6$ cm$^{-3}$.
\par
With this drastic increase in trapping time, and hence trapped molecule number and density, several experimental projects become possible, e.g. the study of state dependent inelastic atom-molecule collision cross sections or the direct observation of vibrational state lifetimes. 
Given sufficient molecular density, collision cross section measurements will determine the viability of molecular evaporative cooling schemes and make the comparison and refinement of theoretical predictions \cite{Ticknor1} possible.
\par
AC-trapping of absolute ground state polar molecules is an essential step on the path to cooling a sample to quantum degeneracy. 
The implementation of the TWIST with 4 rings, rather than 3, allows AC-trapping of molecules in the high field seeking ro-vibrational ground state. Trap depths of  200$\mu$K are achievable, however, we are still investigating the effect of heating due to non-linearities in the field. 
Employing the same methodology as in \cite{Tscherneck1} we expect transfer efficiencies of 30\%-40\% for a single and up to 100\% for a two step STIRAP process from the vibrational states currently trapped in the TWIST to the vibrational ground state.
As the MOTs are overlapped with the TWIST, the remaining atoms can be confined in a magnetic trap after polarization gradient cooling and used to sympathetically cool the trapped molecules further.
\par
In conclusion, we report the successful joining of photoassociative production of ultracold polar molecules with electrostatic trapping.
We have created and trapped ultracold polar NaCs molecules in deeply bound vibrational levels ($\textsl{v}$=19-25) of the $X^{1}\Sigma$
state using this approach. The number and lifetime of the trapped molecules are currently limited by the quality of our vacuum ($4\cdot10^{-9}$
Torr). An improved vacuum will vastly increase the size of our atom clouds 
and hence the production rate of photoassociated molecules as well as the total number and density of trapped molecules.
\par
\begin{acknowledgments}
This work was supported by the National Science Foundation and the The Army Research Office.  Chris Haimberger and Jan Kleinert are grateful to the Laboratory for Laser Energetics for DOE Horton Fellowships.
\end{acknowledgments}

\bibliography{kleinerttwist}

\begin{thebibliography}{35}
\bibitem[{com({\natexlab{b}})}]{comment2}
\bibinfo{note}{The ac electric field limit depends on the frequency, but can be
  as high as the vacuum breakdown limit.}
\end{thebibliography}

\end{document}